\documentclass[review]{elsarticle}
\usepackage{lineno,hyperref}
\usepackage[utf8]{inputenc}

\modulolinenumbers[5]

\journal{Journal of \LaTeX\ Templates}

\usepackage{url}

\usepackage {amsmath}
\usepackage{amssymb}
\usepackage{graphicx}
\usepackage{tabularx}

\begin{document}
\title {Building an Archive with Saada}
\author[oas]{L.Michel}
\author[oas]{C.Motch}
\author[igbmc]{H.N.Nguyen}
\author[oas]{F.X.Pineau}

\address[oas]{Observatoire astronomique de Strasbourg, Université de Strasbourg, CNRS, UMR 7550, 11 rue de l'Université, F-67000 Strasbourg, France}
\address[igbmc]{IGBMC - CNRS UMR 7104 - Inserm U 964
1 rue Laurent Fries / BP 10142 / 67404 Illkirch CEDEX / France }

\begin{abstract}
Saada transforms a set of heterogeneous FITS files or VOTables of various categories (images, tables, spectra \ldots) in a database without writing code. Databases created with Saada come with a rich Web interface and an Application Programming Interface (API). They support the four most common VO services. Such databases can mix various categories of data in multiple collections. They allow a direct access to the original data while providing a homogenous view thanks to an internal data model compatible with the characterization axis defined by the VO.
The data collections can be bound to each other with persistent links making relevant browsing paths and allowing data-mining oriented queries. 

\end{abstract}
\maketitle

\section{Introduction}\label{sec_intro}
Back in 2003, the Survey Science Consortium (SSC) of the XMM-Newton ESA mission was looking for a database system able to store the information extracted from the data files generated by the reduction pipeline.
A specificity of that pipeline is to compute cross correlations between X-ray sources and a set of about 200 archival catalogues and to pack them within the dataset delivered by the SSC to ESA and to the scientific community. The SSC database system developed at Strasbourg has been designed to make the most from these correlation links. It must store the links in a persistent way; it must restore the uniqueness of archival sources correlated several times with different X-ray sources. It must allow the users to filter their selections of X-ray sources either with their intrinsic characteristics or with the characteristics of correlated archival sources. In other terms, the system must support a hierarchical model of linked data. The first solution adopted was based on an object oriented database management system (namely O2) now withdrawn. We then decided to build our own tool based on open software. Considering that these features could be interesting for other teams or missions we decided to offer it to the community. This tool is named Saada (Système Automatique d'Archivage de Données Astronomiques). 
Saada is a tool generating local databases (SaadaDBs)  containing multiple collections of heterogeneous data.
 
The paper starts with an overview of Saada. In section \ref{sec_store} we explain the way heterogeneous data are stored within a SaadaDB. Section \ref{sec_action} adopts a user point of view explaining briefly how to run the software. It is followed by a look inside a SaadaDB.

\section{Saada at a Glance}\label{sec_glance}
The purpose of Saada is to make as easy as possible the implementation of archives containing a large variety of data products.  

The name Saada is related to both the project as a whole and to the database installer. The generic name for a database created by the Saada installer is a SaadaDB which can be later renamed adequately. In this paper, the term Saada refers to the global feature of the project whereas SaadaDB refers to the features of a particular database created with Saada.

A SaadaDB is a standalone database including a storage system, a Web interface and a data loader. It doesn't need to be connected to some remote service. It can be deployed on a laptop or on a big server. It can even work without network. Saada can be used on Linux, MacOS, or MS Windows.

A SaadaDB has the ability to integrate images, spectra, tables or any other files possibly linked to each other. It is managed with either a graphical tool or by scripts and doesn't require writing code. 
The data loader extracts keywords values from input data files (FITS or VOTables) and stores them within a relational database. Data can be seen either as tables of native values or as instances of a data model compliant with the VO. 

Data stored in the SaadaDB can be accessed in different ways:

\begin{enumerate}
\item As a relational database whose tables have been filled with data extracted from the input files by the SaadaDB data loader. This database can be accessed by any application having a connectivity with a relational database management system (RDBMS). 

\item As a VO resource providing a data access through one of the following protocols (SIAP  \cite{2011arXiv1110.0499T}, SSAP \cite{2012arXiv1203.5725T}, CSP or TAP \cite{2011arXiv1110.0497D}). 

\item As a Web application.

\end{enumerate}

\begin{figure}[h]
\includegraphics[width=\textwidth]{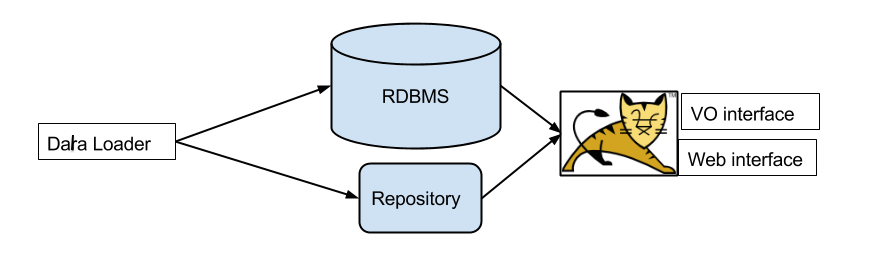}
\caption{SaadaDB modules\label{fig_modules}}
\end{figure}

Figure \ref{fig_modules} sketches the main modules constituting a SaadaDB:

\begin{itemize}
\item The relation database (RDBMS), entirely managed by the SaadaDB, can however host data not managed by that SaadaDB.
\item The repository is a file folder used by the SaadaDB as a storage area.
\item The WEB interface, based on the Saada API, runs as a simple JEE \cite{jee} application (servlets only) hosted by a Tomcat server.
\item VO interfaces are managed by servlets which are actually part of the WEB interface 

\end{itemize}

Data can be queried in a classical way (sky search, keyword filtering) or with more accurate queries as shown in the example below.

\begin{verbatim}
Select all XMM detections  located near ABEL 426
   and having a flux greater than 1e-13
   and correlated with
       at least one Simbad source
       having an obj_type 
            containing the string  ‘Radio’ 
\end{verbatim}

This example uses the computed correlation links mentioned in the introduction to filter data. The exact syntax is detailed in section \ref{sec_saadaql}.

\section{Storing Collections of Heterogeneous Data}\label{sec_store}
\begin{figure}[h]
\includegraphics[width=\textwidth]{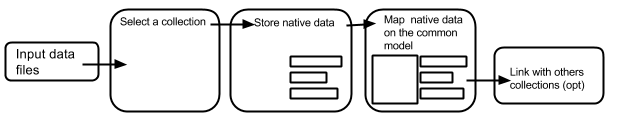}
\caption{Data loading flowchart\label{fig_flowchart}}
\end{figure}

\subsection{Input data files - Data categories}\label{sec_datafile}
The basic job of the data loader consists in extracting keywords values from the input files and to arrange them in the database in a way that makes as easy as possible to reply to user requests (fig \ref{fig_flowchart}).

Saada distinguishes two file categories:
\begin{enumerate}
\item Those from which one can extract data, currently FITS files and VOTables. 

\item The flat-files from which no data is extracted. They are just stored as file references.
\end{enumerate}
Saada data model supports five categories of data whose definition matches the common meaning in astronomy (see table \ref{tab_categories}).  Data files are manually affected to one category. 
As the structure of the input files can be rather complex (multi-extension files) a selection must be done on the set of keywords to be loaded. The keywords of the primary header (FITS) or of the first data table (VOTable) are always taken.
The data-loader can also take the keywords of another FITS extension, which can be either automatically detected or given by the administrator.
By default, the extension loaded is the first matching the specified category.
\begin{table}
\begin{tabularx}{\linewidth}{ c | X X}
    \hline
    Category & Formats & Extension \\ 
    \hline
    FLATILE& Any & No \\
    MISC & FITS, VOTable & Any \\
    SPECTRUM & FITS, VOTable &  Table or image\\
    IMAGE  &  FITS & Image \\ 
    TABLE+ENTRY &FITS, VOTable & Table \\
    \hline
\end{tabularx}
\caption{Product Categories \label{tab_categories}}
\end{table}

For the TABLE category, row values are also stored in a sub category named ENTRY. The row storage works exactly the same way to that of the others categories, taking the column definition instead of the keywords. Pixels or table data are never stored for the others categories. They can however be read to extract the boundaries of the space or energy coverage.

\subsection{Saada Collections}\label{sec_collection}
In a SaadaDB, data are distributed in separate collections created by the administrator. A collection is an abstract container, identified by a name. Collections creation and the choice of the collection where to store a given dataset are completely free. It is a design issue that must be sorted out independently of the products input format. The collection boundaries match the natural scope of the SaadaQL queries (see \ref{sec_saadaql}).

Collections are split in six sub-containers, one per category plus the one for the ENTRY. All have the same twofold internal structure:
\begin{enumerate}
\item The class level storage made of the tables containing the values extracted from the file (except for the FLATFILEs). 

\item The collection level storage containing one table with computed values compliant with the Saada data model.

\end{enumerate}

This data model includes the most relevant parameters in three axis (Space, Time, Energy) of the VO characterization model \cite{2011arXiv1111.2281L} and some observation parameters. It is a subset of the ObsCore model \cite{2011arXiv1111.1758L}. It is always possible to add  by hand  user-defined columns(see \ref{sec_creation}).

All class level tables are joined within the collection level table making the retrieval of all fields from both levels easy.

\subsection{From Heterogeneous Native Data to Homogeneous Mapped Data}\label{sec_hetero}

\subsubsection{Storing Heterogeneous Data at Class Level}\label{sec_classlevel}
The way native data are stored at class level is a bit tricky. By default, data products having the same format are all put in the same class level table. There is one different class level table for each format identified in the set of loaded files.
This rule is actually a bit more flexible since the administrator can force different products to be stored together in the same class. In this case, a class merger adds new columns to the table and when a type conflict occurs (same keywords with different data-type) the type is downcasted  toward the more general one (from boolean to text). The class level tables use a name given by the administrator or by default the name of the first loaded file.
The choice of merging data classes is made by the administrator. This operation mode might be relevant in some cases such as storing in one class data from the same origin (e.g. a reduction software).

\subsubsection{Storing Mapped Data Stored at Collection Level}\label{sec_colllevel}
Mapping class level storage to data model is very challenging because there are rarely one to one links between a keyword and a data model field:
\begin{itemize}
\item The vocabulary used to name a given quantity is very heterogenous;
\item The quantities stored in datafiles are not always complete (e.g. missing unit). The missing data must then be searched in keyword description or in the comments included within the data file or even in the documentation of the software used to generate the product;
\item A data model quantity can result from a computation of multiple values (e.g. applying a threshold on spectrum flux defined with WCS keywords);
\item Relevant information can be spread over multiple extensions or resources.
\end{itemize}
To be accomplished automatically with a good success rate, this operation should be based on a knowledge database identifying the origin of the data product and returning the correct mapping rules.  This issue is likely one of the major justification of the Virtual Observatory.

In order to minimize the risk of mismatch when setting the data model fields, the data mapper of Saada is currently limited to basic identifications.
Data model quantities are either searched among native keywords or given as constant values.  Sometimes they must be formatted  (date) or a unit conversion must be applied (energy range). 

Saada offers two different ways to do this mapping:
\begin{enumerate}
\item The Autodetection mode: The data loader explores the keywords to find out the relevant quantities. It can furthermore interpret WCS keywords or column definitions of a data table or even image pixels. This mode is fine for classical products, but it can lead to irrelevant values when applied to more exotic data.

\item The Mapped mode: The administrator can set a data model field with a constant value or with the name of a keyword from which the value will be taken. If the keyword is not found, the field is no set, but the process continues. 

\end{enumerate}
The two modes are run in parallel with priority level rules set by the administrator. 

\subsection{The Saada Relationships}\label{sec_relation}
In a SaadaDB, data collections can be linked to each other by persistent relationships. A relationship is a named entity joining the data of one given category in a given collection to the data of another category in another collection  (or the same). The links in a relationship can be qualified with numerical values, the qualifiers, which are part of the relationship definition. 
Links are reported on the WEB interface and can be used to refine a data selection. They are also used by the download facility to pack associated data with the selected ones.

\section{Saada in Action}\label{sec_action}

\subsection{Creating a SaadaDB}\label{sec_creation}
A graphical installer achieves the creation of a new SaadaDB. The administrator must set some local resources (location, RDBMS…), a global unit system for the spectral range and a global space frame. Columns can also be added to the collection level model.
All functionalities used by the future SaadaDB are checked at creation time.

\subsection{Managing a SaadaDB}\label{sec_manage}

\begin{figure}[h]
\includegraphics[width=\textwidth]{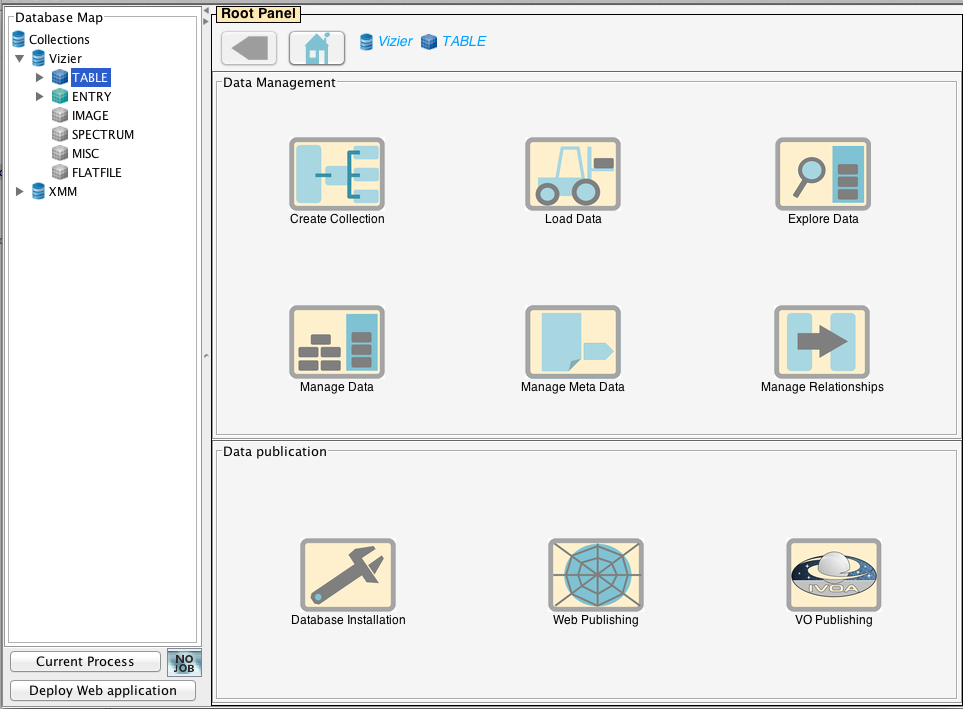}
\caption{Admintool\label{fig_admin}}
\end{figure}

All management operations can be controlled from the graphical tool. The main functionalities are summarized in the table \ref{tab_admin}.

\begin{table}
\begin{tabularx}{\linewidth}{ c | X}
    \hline
    Collection  & Creating, removing and describing collections\\
    Data  & Loading data, removing data, indexing data, editing data-loader filters \\
    Meta data  & Editing UCDs \cite{2011arXiv1110.0525D}, units or description attached to  the data columns. \\
   Relationship  &  Creating, populating, indexing removing relationships.\\
   VO resources & 
Editing VO registry records for specified data collections, creating a TAP service or creating an ObsTap resource.
\\ 
    \hline
\end{tabularx}
\caption{Admintool Features \label{tab_admin}}
\end{table}

Most of these operations can also be run by  \texttt{ant} utility tasks. The XML description of the tasks can be downloaded from the graphical tool.  These scripts allow easy repetition of a management sequence on a wider scale.

\subsection{The Web Interface}\label{sec_web}

\begin{figure}[h]
\includegraphics[width=\textwidth]{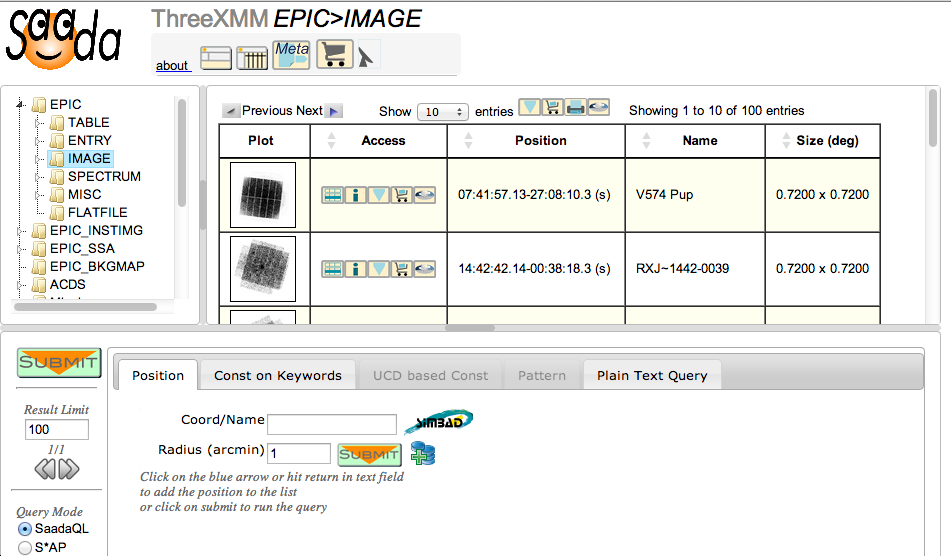}
\caption{Web Interface\label{fig_web}}
\end{figure}

The Web interface \cite{2012ASPC..461..415M} is packed within the distribution. 
It must be deployed (or re-deployed) after the database schema\footnote{Structure of the database: classes and collections in the case of Saada.} has been modified. It can be deployed from the graphical administration tool, or more safely by a script invoking the tomcat deployer.
It is a Rich Internet Application (RIA) built with JQuery and allowing an easy browsing of the database content. Very complex queries can be setup with an editor stacking individual constraints edited within the graphical interface. The final query string can be refined by hand.
The interface proposes a shopping cart feature where the user can store any sort of data in a downloadable ZIP ball. Following a request made by the GBOT project \cite{2013ASPC..475..251B}, the shopping cart content can be forwarded toward a third-party server.  This feature is used to feed a reduction pipeline with data selected from the WEB interface. 
Finally, data can be sent from the Web interface to other VO clients by using a Web profile SAMP connection \cite{2011arXiv1110.0528T}.
 
\subsection{Deploying the VO Services}\label{sec_deploy}
Saada supports four VO protocols served each by one (or more) specific servlet(s). Simple protocols are restricted to their relative Saada category (ENTRY for the cone search, IMAGE for SIAP and SPECTRA for SSA). Publishing a simple service with Saada does not require any action. The service can be connected once the Web interface is deployed just by building proper URLs.

The fourth protocol is TAP. It can endorse any data table.  There is only one TAP service attached to a SaadaDB. The service is set from the graphical administration tool by selecting the data collections to be published. The TAP\_SCHEMA (see this A\&C issue) is automatically updated. Joins between collection storage and class storage tables are taken into account. The TAP implementation uses the CDS/ARI library developed by G. Mantelet \cite{greg}.

\subsection{The Query Language}\label{sec_saadaql}
Saada has its own query language, SaadaQL, where the clause \texttt{SELECT columns FROM table} is replaced with \texttt{SELECT category FROM class IN collection}. The returned columns are not specified since the SaadaQL query always returns a set of identifiers \footnote{Saada relies on an internal identifiers mechanism encoding the location of any record}, which are used by the cache to provide the actual data (see \ref{sec_persistence}).
SQL \texttt{WHERE} statement is replaced with four clauses with each one having its own meaning. They are implicitly ANDed. 

\begin{table}
\caption{SaadaQL Features \label{tab_saadaql-features}}
\begin{tabularx}{\linewidth}{ X | X  }
    \hline
   \texttt{ WherePosition \{\ldots\} } & List of search positions \\ 
   \hline   
   \texttt{ WhereAttributeSaada \{\dots\} } & SQL WHERE statement filtering on collection level fields and on class level fields if the scope of the query is restricted to one class.\\ 
   \hline   
   \texttt{ WhereRelation \{\ldots\} } & List of patterns applied to the vectors formed by the links starting from the queried collection\\ 
   \hline   
   \texttt{ WhereUCD \{\ldots\} } & Filter expressed with Unified Content Descriptor (UCD), a formal vocabulary for astronomical quantities \\ 
    \hline
\end{tabularx}
\end{table}

The query example given in section \ref{sec_glance} is rephrased below with the SaadaQL syntax.

\begin{verbatim}
Select ENTRY From * In CATALOGUE
WherePosition {
  isInCircle("49.94666+41.51305",1,J2000, ICRS)}
WhereAttributeSaada {
   _ep_8_flux > 1e-13}
WhereRelation {
  matchPattern { CatSrcToArchSrc,
        AssObjClass{SimbadEntry},
        AssObjAttSaada{ _obj_type LIKE '%Radio' }}}
\end{verbatim}
SaadaQL is very intricate with the Saada inner data model. It has been designed to be both  concise and expressive to be easily readable on a Web page and modifiable by the user (see \ref{sec_web}). In some extent, the use of SQL statements keeps SaadaQL familiar for SQL or ADQL users.

\section{Inside a SaadaDB}\label{sec_inside}
Basically, a SaadaDB is a Java layer on top of a relational DBMS.  This Java code takes in charge most of the bothering database management (meta data management, data consistency\ldots) and let the users focus on its content. 
The code has been designed in such a way that Saada can operate with as few OS setup as possible. For this reason, Saada does not use Tomcat features such as the connection spooler. 
It actually works with its own spooler well adapted to the specificities of the different supported storage systems.

\subsection{The Storage System}\label{sec_storage}
In theory the Java layer could run with any RDBMS through JDBC \cite{jdbc} but in reality, each system having its own SQL dialect, the application must remain aware to the RDBMS it is connected to. That is why Saada only supports three DBMS:

\begin{enumerate}
\item PostgreSQL (PSQL)  8+: high performances, high scalability;
\item MySQL 5.1+: no real advantage compared to PSQL, but widely used in many places;
\item SQLite: limited concurrency \footnote{Ability of multiple users to access data at the same time} features, but no installation, just a library packed within the Saada distribution.

\end{enumerate}
In order to remain RDBMS agnostic Saada does not use specific extension such as PGSphere for PSQL. The sky search is based on Healpix \cite{2001misk.conf..638O}. All coordinates are stored with their pixel number (level 14 hard coded).  A few SQL procedures are also used.

\subsection{Java Objects and Persistence}\label{sec_persistence}

\begin{figure}[h]
\includegraphics[width=\textwidth]{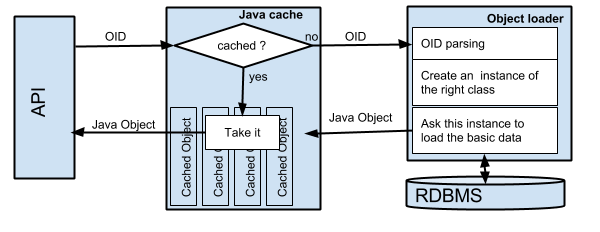}
\caption{Cache\label{fig_cache}}
\end{figure}

Data access is a bit more sophisticated than a simple JDBC callback. It is based on a tight coupling between the SQL storage and the Java classes. Each data record is modeled by a Java class. Basically the Java super-classes models the collection level storage whereas the class level storage is modeled by business classes (sub-classes). Business classes are built on the fly by the data loader and dynamically linked with the application. The instances of the data classes are stored in a cache using the SoftReference mechanism \cite{cache}. This allows the cache to use the whole space available in the memory heap while leaving it available for the garbage collector. When a record is loaded in memory, only the collection level attributes are queried. The class level attributes are all retrieved at the first attempt to access one of them.

The cache mechanism is very efficient to feed the Web interface, which displays on the same page a lot of heterogeneous objects (objects of different classes).  It is however bypassed when the query result must be downloaded in a VOTable or in a ZIP archive. In this case, the data are read in streaming mode and the response is formatted on the fly. 

\subsection{Table Indexation}\label{sec_indexation}
By default, the data loader removes all indexes of the concerned tables before to start and  builds them again after the job is complete. All columns are indexed by default.
This behaviour can be disabled avoiding unnecessary index computation. 
Indexes can be individually managed from the administration tool.

\subsection{Relationship Indexation}\label{sec_rlevel}
When the project started, the SQL system couldn't efficiently process  queries using filters on linked data, especially when they contain numerical predicates associated with a cardinality constraint such as shown by the example below:
\begin{verbatim}
Select all XMM detections 
   correlated with 
      more than 3 Simbad source 
      located at less than 3 arcsec
\end{verbatim}

We took over this limitation by developing a query engine dedicated to the processing of relationship patterns. It  is based on hash maps stored in files and loaded on demand in memory.
%
\subsection{Performance}\label{sec_perf}
The flexibility of Saada prohibits to define performance criteria which are both simple and relevant. A SaadaDB can host a lot of collections with very different sizes. Data collections can host various number of classes, they can be linked with relationships (see \ref{sec_relation}) having a very different number of qualifiers and links. The data ingestion rate depends on the complexity of the collection data mapping.
The global performance of a SaadaDB is also driven by the efficiency of both underlying RDBMS and JDBC driver. 
Taking this into account, the values given below, issued from the experience with the XCatDB \cite{2007ASPC..376..699M}, must be considered as orders of magnitude\footnote{4Core CPU at 2.4GHz, 16GbRAM, PSQL 8.4, Scientific Linux 6}.

\begin{table}
\begin{tabularx}{\linewidth}{ X | c | X}
    \hline
    Catalogue ingestion & 500 to 5000 rows/sec & Depends on the column number. \\    
    Spectrum ingestion & 1 to 5/sec & Depends on the keyword number. \\    
    Images ingestion & 1 to 3/sec & Includes the vignette generation. \\ 
    Flat fules ingestion & few 1000/sec & \\  \hline 
    Source selection 1arcmin around a position & 	100ms & Done on a catalogue of 5,000,000 sources\\
    Sources selection combining filtering on both collection and class level & 2sec & require a join between both  tables (5,000,000 X 1,000,000 rows)\\ 
    Source selection by filtering on relationship links&2sec&The actual relation contains 10,000,000 links\\
    \hline
\end{tabularx}
\caption{Order of magnitude of the performance of a SaadaDB (measured on the XCatDB)}
\end{table}

The volumetry limitation also depends on the RDBMS. We consider that Saada easily supports data collections containing a few tens of millions of entries. Beyhond, some tuning must be done such as  a replication at collection level of the most used class level attributes.

\section{Future and Prospects}\label{sec_future}
After a first beta release \cite{2006ASPC..351...25M}, the concept of Saada has been adapted to the bio-computing paradigm in the Bird project led by Hoan  Nguyen Ngoc \cite{2014IBM}. Both projects are still developed and supported, but separately.

The project evolution is mainly driven by user requests. The versioning is managed in nightly build mode.


Saada has undergone a major evolution in 2014 with the convergence of the inner data model toward the Obscore data model.  The query engine now supports the search by regions. In parallel, a part of the code has been refactored and the website has been redesigned. The foreseen evolutions will be focused on both data loader and VO interfaces.  
The flexibility of the data loader will be improved with the possibility of taking data from more than one FITS extension or one VOTable resource and also with the implementation of arithmetic operations for the mapping between the native data and the collection level data. In collaboration with the CDS, we are  working on the design of the knowledge base mentionned in \ref{sec_colllevel}.
The support of the VO interfaces will also be improved with the implementation of new standards such as SIAP-V2 and Datalink and with the possibility for the administrator to do a fine-tuning on the content of the VO responses.
Looking further ahead, we are thinking about improving the modularity and the flexibility of the WEB interface.

\section{Conclusions}\label{sec_conclusion}
The initial goal of Saada was to pack into a single multiplatform tool two basic functionalities: data management and the WEB interface deployment. The VO interface quickly became the third pillar of the tool. This ambitious way to integrate a large set of features probably contributed to give an image of a product leaving little room for the user setup. That, combined with difficulties encountered by some users to install PostgresQL or MySQL, slowed down the usage of Saada. These matters have been overtaken by the development of the script mode and the support of an embedded database in addition to dozen of other improvements.
Since then, Saada has showed it robustness especially with the XCatDB hosting millions of data files from the XMM-Newton mission. It remains one of the few tools able to transform a set of data files into a database, to publish it in the VO and to provide a Web interface with a few clicks. Among other projects and thanks to its ability to handle heterogeneous data collection, Saada has been chosen by the CDS to host data attached to the Vizier catalogs. 
\section*{References}

\bibliography{SAADA}

\section*{Useful Links}\label{sec_usefullinks}
\begin{enumerate}
\item[] http://saada.unistra.fr
\item[] http://xmmssc-www.star.le.ac.uk/
\item[] http://www.ivoa.net/
\item[] http://ant.apache.org/
\item[] http://www.sqlite.org/
\item[] https://bitbucket.org/xerial/sqlite-jdbc
\item[] http://pgsphere.projects.pgfoundry.org/
\end{enumerate}

\section*{Acknowledgements}
We would like to thank A. Nebot Gomez and F. Gris\'e (Strasbourg Observatory) for reviewing this paper,  Taro L. Saito (Treasure Data, Inc) for developping and supporting the SQLite JDBC driver and all both contributors and users of Saada.

\end {document}